\documentclass[twocolumn,aps,prl,groupedaddress]{revtex4}
\usepackage[latin9]{inputenc}
\usepackage[plainpages=false,pdfpagelabels,colorlinks=true,linkcolor=red,urlcolor=blue,citecolor=blue,pdftitle={Title},pdfauthor={},pdfdisplaydoctitle=true,pdfduplex=DuplexFlipLongEdge]{hyperref}
\usepackage{verbatim}
\usepackage{amsmath,amssymb,bm}
\usepackage{graphicx,xfrac,appendix}
\usepackage{CJK}
\usepackage[english]{babel}  %LatexÖÐµ¥ŽÊ¶ÏÐÐµ«±ÜÃâµ¥žö×ÖÄž±»žô¿ª
\usepackage{subfig}
\usepackage{float}
\makeatletter
\@ifundefined{textcolor}{}
{%
 \definecolor{BLACK}{gray}{0}
 \definecolor{WHITE}{gray}{1}
 \definecolor{RED}{rgb}{1,0,0}
 \definecolor{GREEN}{rgb}{0,1,0}
 \definecolor{BLUE}{rgb}{0,0,1}
 \definecolor{CYAN}{cmyk}{1,0,0,0}
 \definecolor{MAGENTA}{cmyk}{0,1,0,0}
 \definecolor{YELLOW}{cmyk}{0,0,1,0}
}

\usepackage{epstopdf}
\usepackage{xcolor}
\usepackage{tikz}

\makeatother

\begin{document}
\title{Metal-insulator phase transition and topology in a three-component system}
\author{Shujie Cheng and Gao Xianlong}
\address{Department of Physics, Zhejiang Normal University, Jinhua 321004, China}
\date{\today}

\begin{abstract}
 In the framework of the tight binding approximation, we study a non-interacting model on the three-component dice lattice
 with real nearest-neighbor and complex next-nearest-neighbor hopping subjected to $\Lambda$- or V-type
 sublattice potentials. By analyzing the dispersions of corresponding energy bands, we find that the system undergoes a
 metal-insulator transition which can be modulated not only by the Fermi energy but also the tunable extra parameters.
 Furthermore, rich topological phases, including
 the ones with high Hall plateau, are uncovered by calculating the associated band's Chern number. Besides, we also analyze the edge-state spectra and discuss the correspondence between
 Chern numbers and the edge states by the principle of bulk-edge correspondence.\\
\end{abstract}

\maketitle

\section{INTRODUCTION}

Quantum Hall effect is discovered under the condition with a low temperature and strong magnetic fields \cite{Klitzing}.
Since this finding, enormous attention has been poured into the study on the topological properties of the quantum
systems \cite{attention_1,attention_2,attention_3}, particularly on those quantum-engineered systems with quantum anomalous
Hall effect (QAHE) \cite{attention_4,attention_5,attention_6,attention_7,attention_8}, which are free from the extra magnetic fields.

One class of insulators realizing the QAHE breaking the time reversal symmetry \cite{timereversal}, is Chern insulator in condensed matter physics. Their topological properties are directly
characterized by the Hall conductance first proposed by Thouless-Kohmoto-Nightingale-den Nijs (TKNN) in a two-band system \cite{TKNN}. Through TKNN's work, we know that the mathematical concept of Chern number~(C) plays an important role in the topology of Bloch systems. That is, a separated Bloch band corresponds to a well-defined Chern number. In a multiple-band system, the associated Chern number is equal to the quantized Hall
conductance in unit of $e^2/h$ for the lowest occupied band, which determines the topological phases of the system. For instance, the nonzero $C$ is
indicative of a non-trivial phase, whereas $C=0$ is of a topologically trivial phase, demonstrating that the system is a normal insulator. For multiple occupied bands, although the Chern number is not directly related to
the Hall conductance, the summation of the Chern numbers corresponding to these bands finally share the same value of the observable quantized Hall conductance \cite{bulk-edge}.

In the topological classification, Chern insulator belongs to the topological class $A$ \cite{Topo_classify}. In general, it is possible to geometrically engineer a Chern insulator with arbitrary topological index $C$. In addition, higher Hall conductance is accompanied
by larger $C$ with reducing channel resistance in the field of interface transport, possessing more potential for applications.
However, there remains a great challenge to observe the QAHE in real materials \cite{observe_1,observe_2,
 observe_3,observe_4}. Optical-lattice experiments have the advantage of being highly flexible and adjustable, offering a platform for manipulating trapped ultracold atoms to realize the topological non-trivial phases. Accordingly, some attention has been paid to
 the optical-lattice experiments in recent years \cite{attention_4,attention_5,attention_6,attention_7,attention_8,opt_lat_exp_1,
 opt_lat_exp_2,opt_lat_exp_3,opt_lat_exp_4,opt_lat_exp_5,opt_lat_exp_6,opt_lat_exp_7,opt_lat_exp_8,opt_lat_exp_9,opt_lat_exp_10,
 opt_lat_exp_11}. With the benefits of quite a few efforts \cite{attention_4,attention_5,attention_6,attention_7,attention_8,
 opt_lat_exp_4,opt_lat_exp_7,effort_1,effort_2,effort_3}, two quintessential models with rich topological phenomena, i.e.,
 Haldane model \cite{opt_lat_exp_8,Haldane_model} and  Haper-Hofstadter model \cite{opt_lat_exp_9,Hofstadter_model} have been
 well implemented. These mature techniques have had a significant impact on subsequent researches \cite{sub_research_1,sub_research_2,
 sub_research_3,sub_research_4,sub_research_5,sub_research_6,sub_research_7} and will increase the feasibility to realize the
 higher Hall conductance or larger Chern number with QAHE.

 In this paper, we focus on a non-interacting model on the three-component dice lattice \cite{dice_lattice_1,dice_lattice_2,
 dice_lattice_3,dice_lattice_4,dice_lattice_5,dice_lattice_6,dice_lattice_7,dice_model_1,dice_model_2,dice_model_3} with real nearest neighbor
 and complex next-nearest neighbor hopping under $\Lambda$- or V-shaped sublattice potentials. We focus on
 the $1/3$ filling and $2/3$ filling cases, corresponding to the lowest band occupied and lower two
 bands occupied, respectively. By analyzing the band structures, we find that the system will have two kinds of phases, metallic phase and bulk
 insulating phase, thus, experiencing a metal-insulator phase transition. The metal-insulator phase diagram is plotted in the $\Delta$-$\phi$ parameter space, with $\Delta$ the strength of
 the tunable on-site potentials and $\phi$ the phase of the next-nearest-neighbor hopping. Further, we investigate the topological properties in the bulk insulating phase. Results suggest that there exist non-trivial topological
 phases with Hall conductance equal to $\pm 1$, $\pm 2$, in units of $e^2/h$. Furthermore, the quantization of the Hall conductance
 can be seen from the edge-state energy spectra and the large Chern numbers are self-consistently analyzed by the principle
 of bulk-edge correspondence \cite{principle_of_bulk_edge}.

 The paper proceeds as follows. Section \ref{S2} describes the model and its Hamiltonian in both the real and momentum space.
 Section \ref{S3} contains the analysis of the band structures, the topological phases, as well as the edge-state spectra
 and Chern numbers. Section \ref{S4} summarizes the results of these investigations.

\section{\label{S2}MODEL AND HAMILTONIAN}
\begin{figure}[H]
  \centering
  \includegraphics[width=0.5\textwidth]{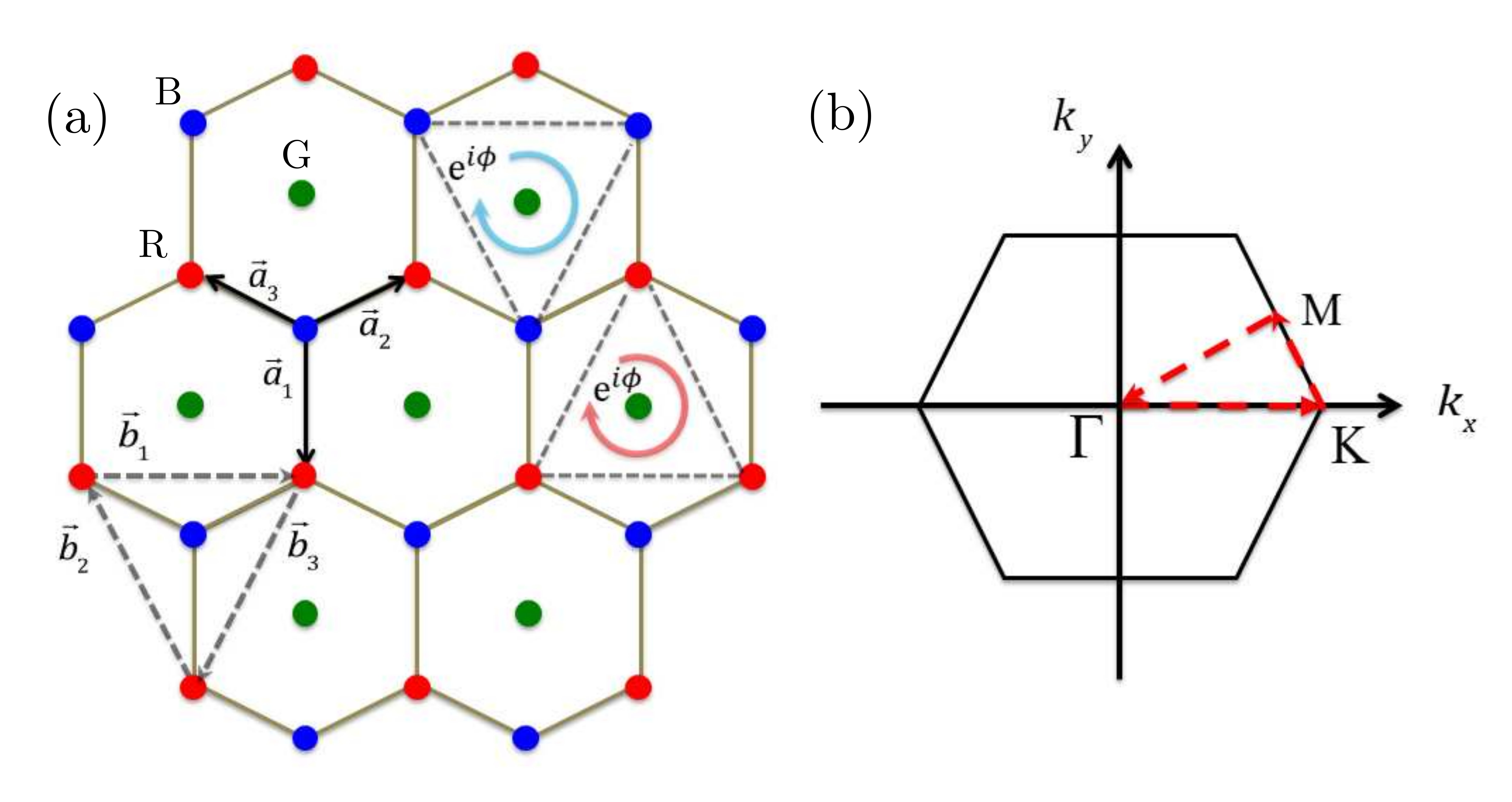}
  \caption{(a) The schematic diagram of the three-band $\Lambda$- or V-type dice model. The red, blue and green
   dots denote the R, B, G sublattice sites, respectively. The vectors $\bm{a}_n$ ($n=1,2,3$) connect the nearest
  neighbors which belong to different sublattices. The vectors $\bm{b}_n$ ($n=1,2,3$) connect
  the nearest neighbors which belong to the same R and B sublattice sites. The circle arrows show the direction of hopping
  with a phase factor $\phi$ in $e^{i\phi}$. (b) The first Brillouin zone. $\Gamma$, K, and M are high-symmetry points which are connected
  by three red dashed arrows. The distance between adjacent sites has been set
  as unit length.}\label{f1}
 \end{figure}

 In this paper, we consider a non-interacting model based on the dice lattice, shown in Fig.~\ref{f1}, with three interpenetrating triangle sublattices [denoted by R (red dots), B (blue dots) and G (green dots)]. The single-particle
 Hamiltonian of the model consists of three terms:
 \begin{equation}\label{eq1}
 \hat{H}=\hat{H}_1+\hat{H}_2+\hat{H}_3.
 \end{equation}

 The first term $\hat{H}_{1}$ describes the hopping between adjacent sites which belong to different sublattices and is
 expressed as
 \begin{equation}\label{eq2}
 \begin{aligned}
 \hat{H}_1&=\sum_{\langle\mathbf{R}_i,\mathbf{B}_j\rangle}t\left(\hat{c}^\dag_{\mathbf{R}_i}\hat{c}_{\mathbf{B}_j}+H.c.\right)
 +\sum_{\langle\mathbf{R}_i,\mathbf{G}_\ell\rangle}t_1\left(\hat{c}^\dag_{\mathbf{R}_i}\hat{c}_{\mathbf{G}_\ell}+H.c.\right)\\
 &+\sum_{\langle\mathbf{B}_j,\mathbf{G}_\ell\rangle}t_1\left(\hat{c}^\dag_{\mathbf{B}_j}\hat{c}_{\mathbf{G}_\ell}+H.c.\right),
 \end{aligned}
 \end{equation}
 where $t$ is the hopping amplitude between adjacent R sites and B sites, $t_1$ is the real hopping amplitude between
 adjacent R (or B) and G sites, $\hat{c}_{\mathbf{R}_i}, \hat{c}_{\mathbf{B}_j}$, and $\hat{c}_{\mathbf{G}_\ell}$ are the corresponding
 fermionic annihilation operators defined on the relevant sites $\mathbf{R}_i$, $\mathbf{B}_j$, and $\mathbf{G}_\ell$ of sublattice
 R, B, and G, respectively. $\langle\cdots\rangle$ means the nearest-neighbor hopping.

 Furthermore, we consider the hopping between two nearest neighbor sites only in R and B sublattice site. Then the
 Hamiltonian $\hat{H}_2$ is written as
 \begin{equation}\label{eq3}
 \begin{aligned}
 \hat{H}_2&=\sum_{\langle\mathbf{R}_i,\mathbf{R}_j\rangle}\left(t_2e^{i\phi}
 \hat{c}^\dag_{\mathbf{R}_i}\hat{c}_{\mathbf{R}_j}+H.c.\right)\\
 &+\sum_{\langle\mathbf{B}_i,\mathbf{B}_j\rangle}\left(t_2e^{i\phi}
 \hat{c}^\dag_{\mathbf{B}_i}\hat{c}_{\mathbf{B}_j}+H.c.\right),
 \end{aligned}
 \end{equation}
 where $t_2e^{\pm i\phi}$ is the hopping amplitude, $\pm$ correspond to the clockwise (anticlockwise) direction of the
 hopping. $t_2$ is a real number and $\phi$ denotes the phase. These flux configurations can be obtained by the artificial
 gauge field \cite{attention_4,attention_5,attention_6,attention_7,attention_8,opt_lat_exp_1,opt_lat_exp_2,opt_lat_exp_3}.

 The final term, $\hat{H}_{3}$, describes an on-site potential of the form \cite{V-type}
  \begin{equation}\label{eq4}
  \hat{H}_3=\gamma_{1}\Delta\sum_{\mathbf{R}_i}\hat{c}^\dag_{\mathbf{R}_i}\hat{c}_{\mathbf{R}_i}+\gamma_{1}\Delta\sum_{\mathbf{B}_i}
  \hat{c}^\dag_{\mathbf{B}_i}\hat{c}_{\mathbf{B}_i}+\gamma_{2}\Delta\sum_{\mathbf{G}_i}\hat{c}^\dag_{\mathbf{G}_i}\hat{c}_{\mathbf{G}_i},
  \end{equation}
 where $\Delta$ is the tunable parameter of the potential and $\gamma_{1}~(\gamma_{2})$ is modulation slope of $\Delta$.
 This shape of the potential can be viewed as a three-band model in
 the lattice case of three discrete levels forming a $\Lambda$- ($\gamma_{1}<\gamma_{2}$) or V-type ($\gamma_{1}>\gamma_{2}$) structure. Such on-site potentials are a bit different from previous works on the dice model \cite{dice_model_1,dice_model_2,
 dice_model_3} and can also be realized by tuning single-beam lattice depths \cite{opt_lat_exp_1,opt_lat_exp_4,opt_lat_exp_8,
 opt_lat_exp_9}.

 In our research, we consider a sufficiently large system such that the translational symmetry is preserved. Then we
 map the single-particle Hamiltonian $\hat{H}$ onto the momentum space \cite{su3_1,su3_2,su3_3}. The Bloch Hamiltonian is written as
 \begin{equation}\label{eq5}
 \hat{\mathcal{H}}(\mathbf{k})=I(\mathbf{k})+\mathbf{d}(\mathbf{k})\cdot\vec{\lambda},
 \end{equation}
 where $I(\mathbf{k})$ is a scalar leading to a overall shift of three eigenvalues of $\hat{\mathcal{H}}(\mathbf{k})$,
 $\mathbf{d}(\mathbf{k})$ is an eight-dimensional real vector, and $\vec{\lambda}$ is a vector of Gell-Mann matrices \cite{Gell}.
 In fact, the Chern number will not be influenced by the scalar $I(\mathbf{k})$, and is only determined by the coefficient
 vectors $\mathbf{d}(\mathbf{k})$. The discrete Fourier transformation is performed in the three-component basis
 $\left(\hat{c}_{\mathbf{k},{\rm R}},\hat{c}_{\mathbf{k},{\rm B}},\hat{c}_{\mathbf{k},{\rm G}}\right)^T$, with
 \begin{equation}\label{eq6}
 \begin{aligned}
 \hat{c}_{\mathbf{k},{\rm R}}&=\frac{1}{\sqrt{N}}\sum_{\mathbf{R}_j}
 e^{-i\mathbf{k}\cdot\mathbf{R}_j}\hat{c}_{\mathbf{R}_j},\\
 \hat{c}_{\mathbf{k},{\rm B}}&=\frac{1}{\sqrt{N}}\sum_{\mathbf{B}_j}
 e^{-i\mathbf{k}\cdot\mathbf{B}_j}\hat{c}_{\mathbf{B}_j},\\
 \hat{c}_{\mathbf{k},{\rm G}}&=\frac{1}{\sqrt{N}}\sum_{\mathbf{G}_j}
 e^{-i\mathbf{k}\cdot\mathbf{G}_j}\hat{c}_{\mathbf{G}_j},
 \end{aligned}
 \end{equation}
 where $N$ is the number of the cells. The components of the vector $\bf{d}(\bf{k})$ are obtained as follows,
 \begin{equation}\label{eq7}
 \begin{aligned}
 d_1&=t\sum_{s}\cos\left(\mathbf{k}\cdot\bm{a}_s\right),~~d_2=t\sum_{s}\sin\left(\mathbf{k}\cdot\bm{a}_s\right),\\
 d_4&=d_6=t_1\sum_{s}\cos\left(\mathbf{k}\cdot\bm{a}_s\right),\\
 d_7&=-d_5=t_1\sum_{s}\sin\left(\mathbf{k}\cdot\bm{a}_s\right),\\
 d_3&=-2t_2\sin{\phi}\sum_{s}\sin(\mathbf{k}\cdot\bm{b}_s),\\
 d_8&=\frac{\gamma_{1}-\gamma_{2}}{\sqrt{3}}\Delta+\frac{2t_2}{\sqrt{3}}\cos{\phi}\sum_{s}\cos(\mathbf{k}\cdot\bm{b}_s),
 \end{aligned}
 \end{equation}
 where the six vectors $\bm{a}_n$ and $\bm{b}_n$ ($n=1,2,3$), shown in Fig.~\ref{f1}(a), are expressed as
 \begin{equation}\label{eq8}
 \begin{aligned}
 \bm{a}_1&=\binom{0}{-1},\quad\bm{a}_2=\frac{1}{2}\binom{\sqrt{3}}{1},
 \quad\bm{a}_3=\frac{1}{2}\binom{-\sqrt{3}}{1},\\
 \bm{b}_1&=\binom{\sqrt{3}}{0},\quad\bm{b}_2=\frac{1}{2}\binom{-\sqrt{3}}{3},
 \quad\bm{b}_3=-\frac{1}{2}\binom{\sqrt{3}}{3}.
 \end{aligned}
 \end{equation}

In the next section, we will study the properties of the band structures and topology of the system under $1/3$ and $2/3$ fillings with the lowest and lower two bands occupied, respectively. We set $t=1$ as the unit of energy, and choose $t_1=0.5$, $t_2=0.526$, $\gamma_1=5$, and $\gamma_2=7$.

\section{\label{S3}RESULTS and DISCUSSIONS}

 \subsection{Band structures}

 \begin{figure}[H]
  \centering
  \includegraphics[width=0.5\textwidth]{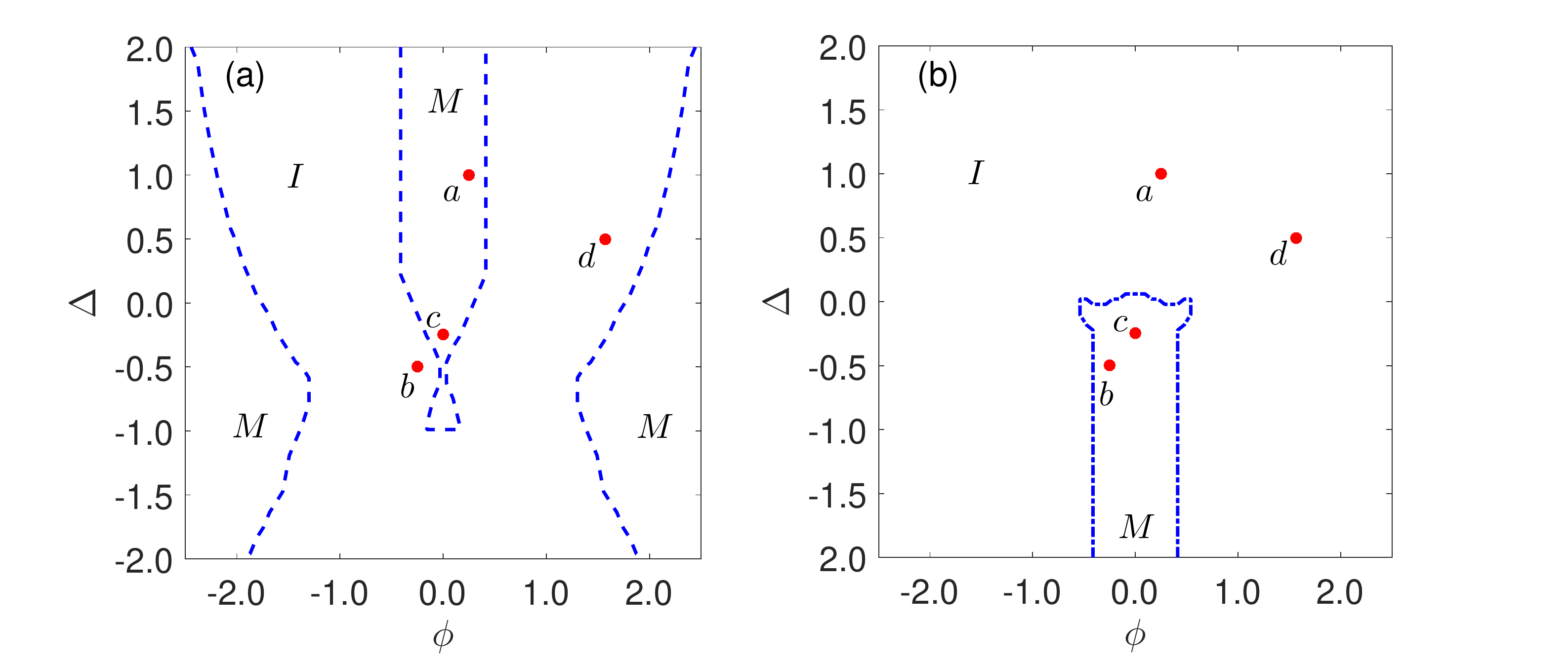}
  \caption{$\Delta$-$\phi$ phase diagram for the case of $1/3$ filling (a) and $2/3$ filling (b). The metallic phase $(M)$ is separated from the bulk insulating phase (I).
  In each case, we choose four typical parameter points $a$, $b$, $c$, and $d$ discussed in details in the main text.
  For 1/3 filling, system is in the metallic phase at the $a$ point, while it is in the bulk insulating phase at the $b$ and $d$ points. The opposite is obtained for the ones at 2/3 filling. For the $c$ ($d$) point, the system is always in the metallic (insulating) phase, no matter what the filling is.
  }\label{f2}
 \end{figure}

 First, we numerically investigate the band structures of this system by analysing the dispersions of three bands. The first
 Brillouin zone is shown in Fig.~\ref{f1}(b) and the $\Gamma$, K, and M are high-symmetry points \cite{Kpoints}. By diagonalizing
 the Hamiltonian $\hat{\mathcal{H}}(\mathbf{k})$ with the known components of $\mathbf{d(k)}$ in Eq.~(\ref{eq7}), we obtain the energy dispersions along the high-symmetry $k$-points along the path $\Gamma$-K-M-$\Gamma$ at a chosen parameter point $(\phi,~\Delta)$,
 where $\phi$ has been converted to radians.

 \begin{figure}[H]
  \centering
  \includegraphics[width=0.5\textwidth]{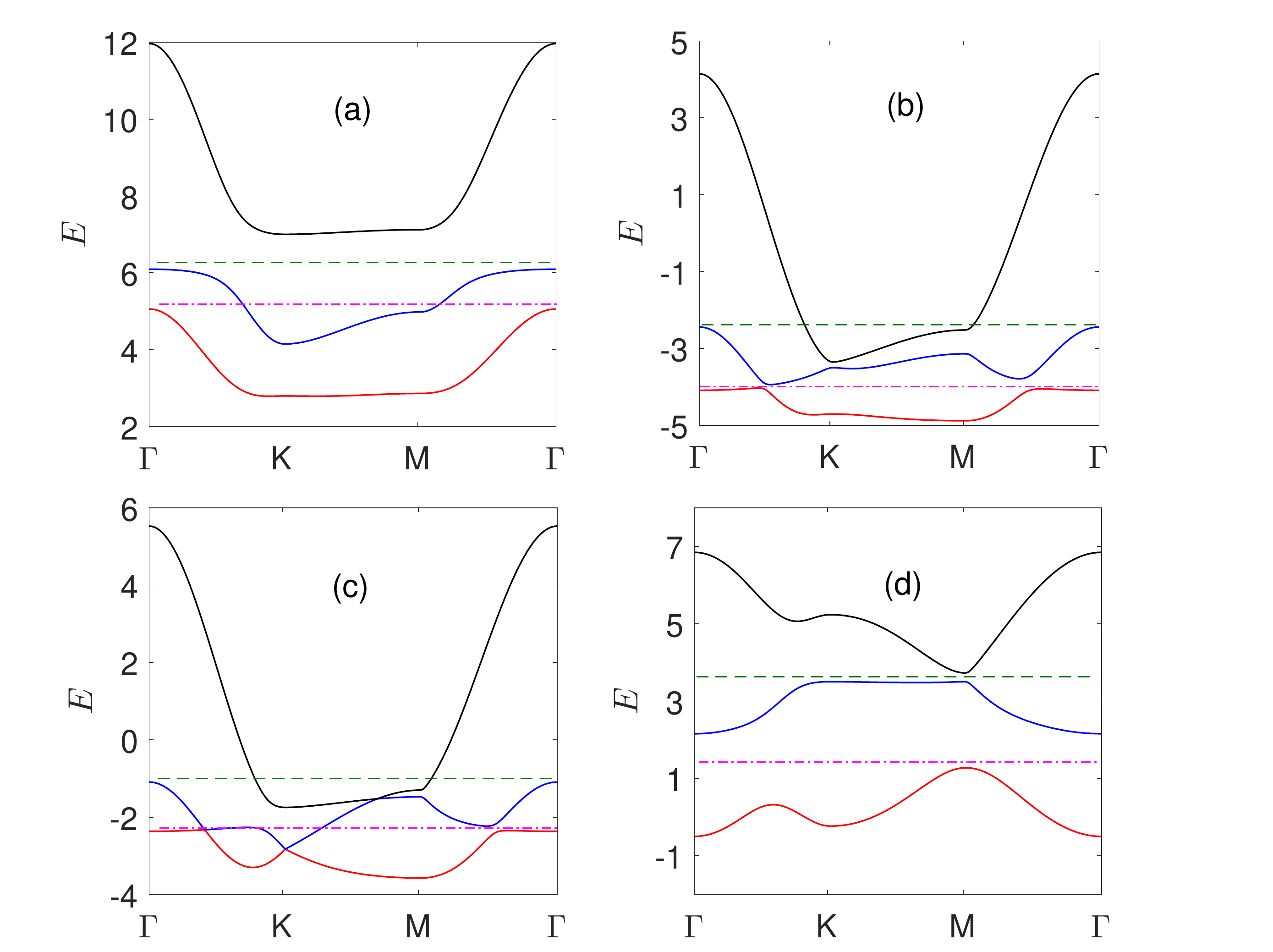}
  \caption{Dispersion relations of the three bands along the high-symmetry $k$-points along the path $\Gamma$-K-M-$\Gamma$. The red, blue, and black lines
  correspond to the dispersion relations of the three bands from the bottom to the top, respectively. (a) $\phi_a=0.25$, $\Delta_a=1$;
  (b) $\phi_b=-0.25$, $\Delta_b=-0.5$; (c) $\phi_c=0$, $\Delta_c=-0.25$; (d) $\phi_d=\pi/2$, $\Delta_d=0.5$. The lower Fermi energy
  (magenta dot-dashed line) and higher Fermi energy (green dashed line) correspond to the case of $1/3$ and $2/3$ filling, respectively.
  }\label{f3}
 \end{figure}

Based on the relative position of the valance band and the conduction band, we notice that there are two different
 phases in this model, the metallic phase and bulk insulating phase. To be precise, when the top of the valance band is
 higher than the bottom of the conduction band, the system is in the metallic phase ($M$). Otherwise, the system is in
 the bulk insulating phase ($I$). According to this feature, the metal-insulator phase diagrams
 are shown in Figs.~\ref{f2}(a) and \ref{f2}(b), corresponding to $1/3$ filling and $2/3$ filling, respectively.
These phases are symmetrically distributed with respect to $\phi$. Four typical parameter points, $a$, $b$, $c$ and $d$ (red dots), are chosen in each of the $\Delta$-$\phi$ phase diagram. For 1/3 filling case, system is in the metallic phase at the $a$ point, while
 it is in the bulk insulating phase at the $b$ and $d$ points. The situations are opposite for the ones for 2/3 filling. For the $c$ ($d$)
 point, the system remains in the metallic (insulating) phase, no matter what the filling is. One can adjust the Fermi energy $E_{f}$ or the parameters of $\Delta$ and $\phi$ to
 change the system from the metallic to bulk insulating phases. For instance, if the system is initially in the metallic phase ($a$ point),
 we can increase $E_{f}$ by adding the filling up to $2/3$, or change the parameters up to the $b$ ($d$) point, to achieve the bulk insulating phase.

 In order to comprehend these two phases intuitively, we plot the dispersion relations of the three bands at the four chosen
 parameter points $(\phi_a,~\Delta_a)$, $(\phi_b,~\Delta_b)$, $(\phi_c,~\Delta_c)$ and $(\phi_d,~\Delta_d)$, shown in Figs.~\ref{f3}(a), \ref{f3}(b), \ref{f3}(c), and \ref{f3}(d), respectively. In each diagram, we select two Fermi energies for
 reference, corresponding to the case of $1/3$ filling (magenta dot-dashed line) and $2/3$ filling (green dashed line), respectively. The red, blue, and black solid lines represent the dispersions of the bands ranging from the bottom to the top, respectively. In Fig.~\ref{f3}(a), when the $E_{f}$ is chosen at $1/3$ filling, the fully occupied bottom band and a partially
 occupied middle band lead to a metallic phase, although
 the middle and the bottom band avoid touching each other. When the
 $E_{f}$ is tuned up to $2/3$ filling, there is a significant band gap between the top and the middle band, leading to a completely empty conduction and a fully occupied valence band. Therefore, the system is in the
 bulk insulating phase for $2/3$ filling. With the same reasons, it is not difficult to interpret that the system
 is in the bulk insulating phase at $1/3$ filling, and in the metallic phase at $2/3$ filling for the case shown in Fig.~\ref{f3}(b).
 Moreover, the system is always in the metallic phase shown in Fig.~\ref{f3}(c) and in the bulk insulating phase shown in Fig.~\ref{f3}(d), respectively, no matter what the filling is. All these are manifested in the metal-insulator phase
 diagrams in Fig.~\ref{f2}. In the following, we will further investigate the topological properties in the $\Delta$-$\phi$ phase diagram.

\subsection{Topological phases}
Motivated by the Haldane model \cite{Haldane_model} and the
 dice models \cite{dice_model_1,dice_model_2,dice_model_3}, we try to understand whether there are any topological
 properties in the bulk insulating phase or not. For the three-band system considered, when $E_{f}$ is within a band gap,
 the Hall conductance \cite{TKNN} can be defined as
 \begin{equation}
 \sigma_{H}(E_{f})=\frac{e^2}{h}\sum_{E_{n}<E_{f}}C_{n},
 \end{equation}
 where $C_{n}$ is Chern number of the $n$th fully occupied band and is expressed as
 \begin{equation}
 C_n=\frac{1}{2\pi}\oint_{\partial{BZ}}\mathbf{A}_n(\mathbf{k})\cdot d\mathbf{k},
 \end{equation}
 where $n~\epsilon~\{1,2,3\}$ is the band index and its ascending order corresponds to three energy bands from bottom
 to top; $\partial{BZ}$ is the boundary of the first Brillouin zone; $\mathbf{A}_{n}$ is the Berry connection with
 $\mathbf{A}_n=-i\langle\psi_n(\mathbf{k})|\mathbf{\nabla_{k}}|\psi_n(\mathbf{k})\rangle$ and $|\psi_n(\mathbf{k})\rangle$
 is the corresponding eigenvector. Without loss of generality, we use two quantities, $C_{\frac{1}{3}}$ and $C_{\frac{2}{3}}$,
 to character the topological properties under $1/3$ and $2/3$ filling, respectively. The relationship between the topological number of the filling (
 $C_{\frac{1}{3}}$, $C_{\frac{2}{3}}$) and band Chern numbers ($C_{1}$ and $C_{2}$) is that $C_{\frac{1}{3}}=C_{1}$
 and $C_{\frac{2}{3}}=C_{1}+C_{2}$.

\begin{figure}[H]
  \centering
  \includegraphics[width=0.5\textwidth]{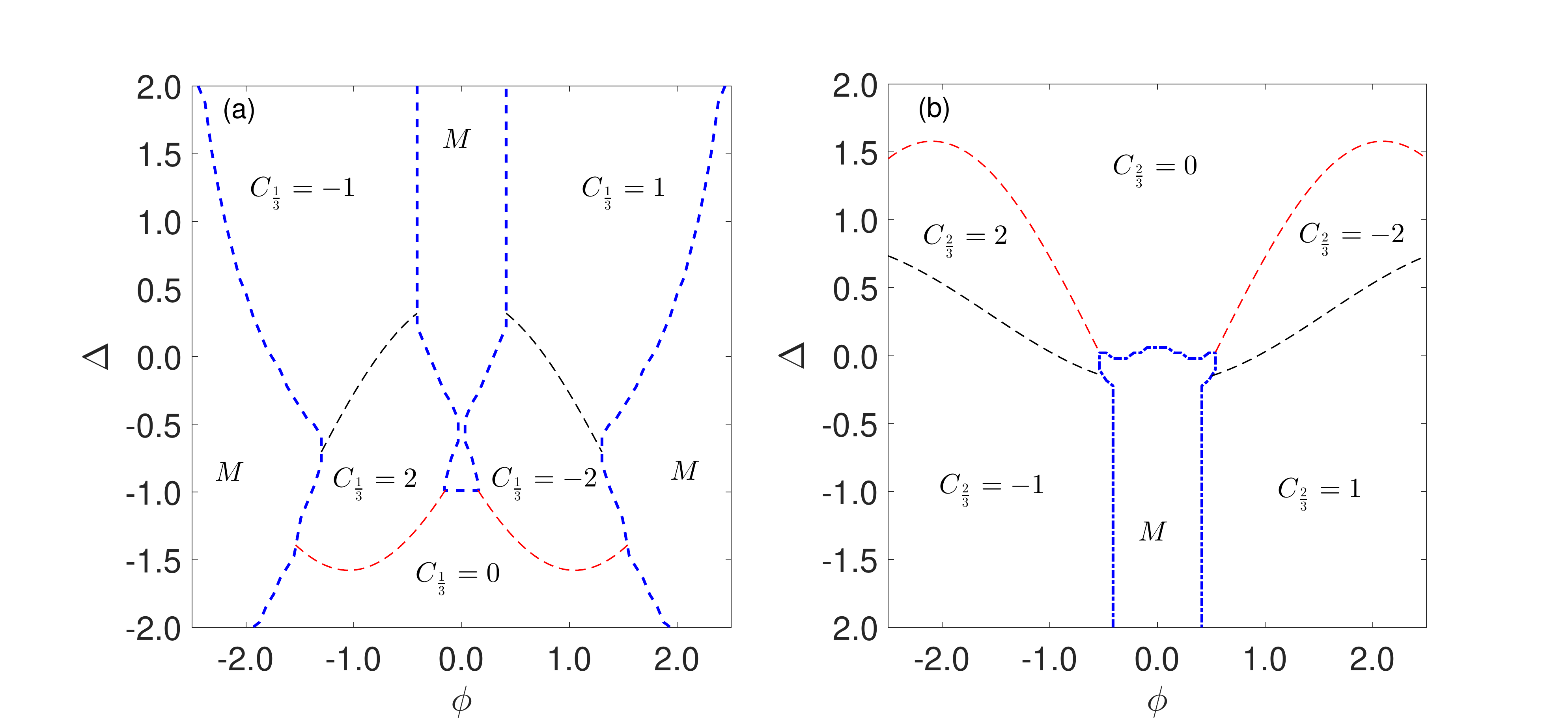}\\
  \caption{The topological $\Delta-\phi$ phase diagram for the (a) $1/3$ filling and (b) $2/3$ filling, respectively.
  The Chern numbers have been marked in these two diagrams. Except for the metallic phase ($M$) surrounded by the blue dashed lines intersected with the coordinate axes,
   the bulk insulating phase is divided into several
  regions by red dashed lines which distinguish the topological non-trivial phase from the topological trivial one,
  and by black dashed lines which separate the topological non-trivial phases with different Chern numbers.}\label{f4}
 \end{figure}

The topological $\Delta$-$\phi$ phase diagram is obtained shown in Figs.~\ref{f4}(a) and \ref{f4}(b) by calculating the $C_{\frac{1}{3}}$ and $C_{\frac{2}{3}}$, corresponding to the topological numbers for the $1/3$ and $2/3$ filling of the system, respectively. In Fig.~\ref{f4}, there are
several phase boundary lines shown in red and black dashed lines in the bulk insulating phase, accompanied by energy band closing \cite{Haldane_model,band_closing}. In fact, the band crossing lines also appear in the metallic phase without changing the intrinsic properties being a metal. Particularly, in each diagram, the topological non-trivial phase is separated from the topological trivial one by the red dashed lines, whereas the topological non-trivial phases are separated from the different nonzero Chern numbers by the black dashed lines.

From the phase diagrams, we can see intuitively that there are abundant quantum phases in the system. In Fig.~\ref{f4}(a), besides the metallic phase regions, there are topological non-trivial phases with
 $C_{\frac{1}{3}}=\pm 1$ and $C_{\frac{1}{3}}=\pm 2$, as well as the topological trivial phase with $C_{\frac{1}{3}}=0$.
Moreover, when we tune the parameters continuously, the system goes through rich different phases. For instance,
 when $\Delta=-0.25$, by increasing of $\phi$, the system can circularly undergo with six phases:
 \begin{equation*}\label{p1}
%\left(
  \begin{array}{ccc}
    C_{\frac{1}{3}}=-1 & \Rightarrow & C_{\frac{1}{3}}=+2 \\
    \Uparrow &   & \Downarrow \\
    M &   & M\\
    \Uparrow &   & \Downarrow \\
    C_{\frac{1}{3}}=+1 & \Leftarrow & C_{\frac{1}{3}}=-2, \\
  \end{array}
%\right)
 \end{equation*}
 where $M$ is short for the metallic phase.
 When $\Delta=0.5$, system will cycle through five phase regions with increasing $\phi$:
 \begin{equation*}\label{p2}
 \begin{array}{ccccc}
   C_{\frac{2}{3}}=-1 & \Rightarrow & C_{\frac{2}{3}}=+2 & \Rightarrow & C_{\frac{2}{3}}=0 \\
   \Uparrow &  &  &  & \Downarrow\\
    C_{\frac{2}{3}}=+1 &   & \Longleftarrow  &   & C_{\frac{2}{3}}=-2. \\
 \end{array}
 \end{equation*}
Similarly, for fixing $\phi$, the system experiences with rich phases when we change the tunable parameter $\Delta$ of the potential.

 \subsection{Edge states and Chern numbers}
 \begin{figure}[H]  %[htbp]ÖÐµÄhÊÇž¡¶¯µÄÒâËŒ
    \centering    %ŸÓÖÐ

    \subfloat[zigzag edge] %µÚÒ»ÕÅ×ÓÍŒ
    {
        \begin{minipage}[t]{0.5\textwidth}
            \centering          %×ÓÍŒŸÓÖÐ
            \includegraphics[width=0.9\textwidth]{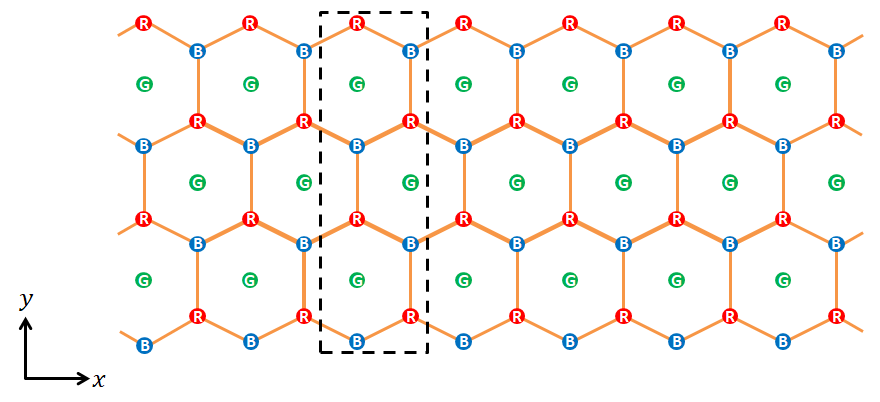}   %ÒÔÐÐ¿íµÄ0.5±¶ŽóÐ¡ÏÔÊŸ
        \end{minipage}%
    }%×¢ÒâÕâÀï²»ÄÜ»Ø³µ¿ÕÐÐ£¬·ñÔòÁœÕÅÍŒ»áÉÏÏÂÅÅÁÐ£¬¶ø²»ÊÇ²¢ÅÅÅÅÁÐ

    \subfloat[armchair edge] %µÚ¶þÕÅ×ÓÍŒ
    {
        \begin{minipage}[t]{0.5\textwidth}
            \centering      %×ÓÍŒŸÓÖÐ
            \includegraphics[width=0.9\textwidth]{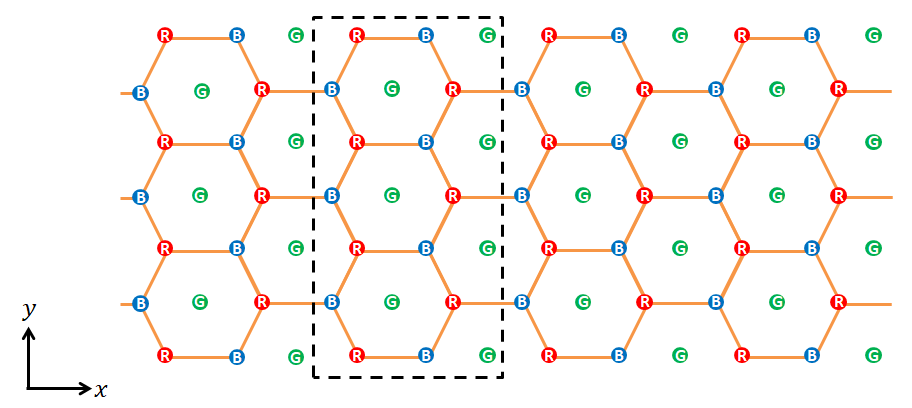}   %ÒÔÐÐ¿íµÄ0.5±¶ŽóÐ¡ÏÔÊŸ
        \end{minipage}
    }%

    \caption{Two schematic diagrams of the system with a periodic boundary condition in the $x$ direction and an open
    boundary condition in the $y$ direction. Two types of boundaries are studied: (a) zig-zag edge; (b) armchair egde. The lattice structures surrounded by the black dotted lines represent
    the periodic repeating units which contain $N_{zigzag}$ and $N_{armchair}$ lattice sites, respectively.} %  % ŽóÍŒÃû³Æ
    \label{f5}  %ÍŒÆ¬ÒýÓÃ±êŒÇ
\end{figure}

 According to the research in Ref. \cite{bulk-edge}, we know that the quantization of the Hall conductance or the total
 Chern number can be readily seen from the edge-state spectrum in static systems, known as the bulk-edge correspondence. In
 this subsection, we study the bulk-edge correspondence by considering a cylindrical geometry with a periodic boundary condition
 in the $x$ direction and an open boundary condition in the $y$ direction. Two types of lattice
 geometries with zigzag and armchair edge are studied, shown in Figs.~\ref{f5}(a) and \ref{f5}(b), respectively, surrounded by the dashed boxes representing
 the periodic repeating units, each of them containing $N_{zigzag}$ and $N_{armchair}$ lattice sites. The Hamiltonian
 $\hat{H}(k_{x})$ used for calculating the energy spectrum $E_{k_{x}}$ can be treated as a function of the good quantum number
 $k_{x}$.
\begin{figure}[H]
  \centering
  \includegraphics[width=0.5\textwidth]{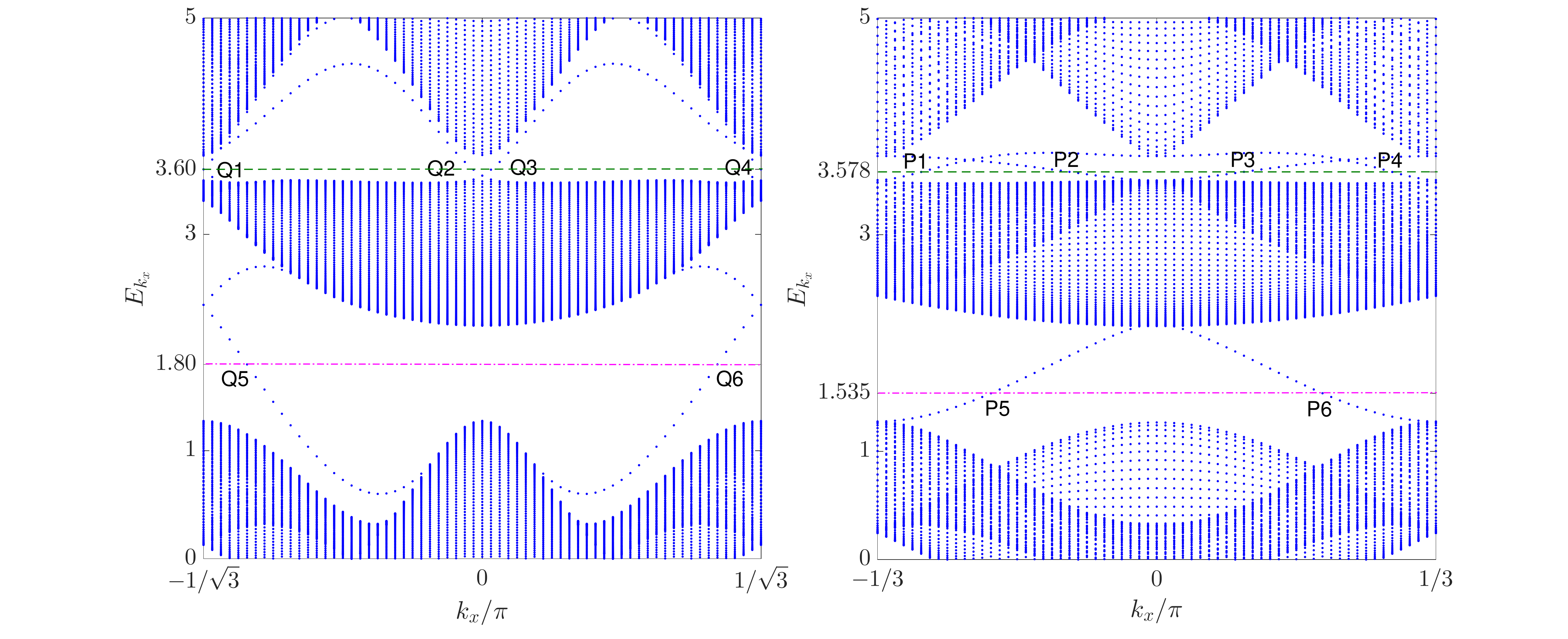}\\
  \caption{ Two edge-state spectra of the system with a periodic boundary condition in the $x$ direction and an open boundary
  condition in the $y$ direction at the parameter point $(\phi_{d},~\Delta_{d})=(\pi/2,~0.5)$. Magenta
  dot-dashed and green dashed lines denote the chosen $E_{f}$ for $1/3$ filling and $2/3$ filling, respectively. (a) Edge-state spectra
  for the zigzag-edge case. Q5 and Q6 are a pair of edge mode at $E_{f}=1.80$ for 1/3 filling; Q1 and Q4, Q2
  and Q3 are two pairs of edge modes at $E_{f}=3.60$ for 2/3 filling. (b) Edge-state spectra for the armchair-edge case. P5 and P6 are a pair of chiral edge mode at $E_{f}=1.535$ for 1/3 filling; P1 and P4, P2 and P3
  are two pairs of edge modes at $E_{f}=3.578$ for 2/3 filling. The results indicates that, no matter what kind of edge the system has,
  there is a pair of edge modes at $1/3$ filling corresponding to $C_{\frac{1}{3}}=1$, and two pairs of edge modes at $2/3$
  filling corresponding to $C_{\frac{2}{3}}=-2$.}\label{f6}
 \end{figure}

By choosing $N_{zigzag}=299$ and $N_{armchair}=243$, two associated edge-state spectra at the parameter point $(\phi_{d},~\Delta_{d})=(\pi/2,~0.5)$ are calculated, shown in Figs.~\ref{f6}(a) and \ref{f6}(b), respectively. Intuitively,
there is a pair of edge modes at $1/3$ filling corresponding to a topological non-trivial phase with $C_{\frac{1}{3}}=1$, and
 two pairs of edge modes at $2/3$ filling corresponding to a topological non-trivial phase with $C_{\frac{2}{3}}=-2$ with no dependence on the edge of the system.
 \begin{figure*}[t]
  \centering
  \includegraphics[width=\textwidth]{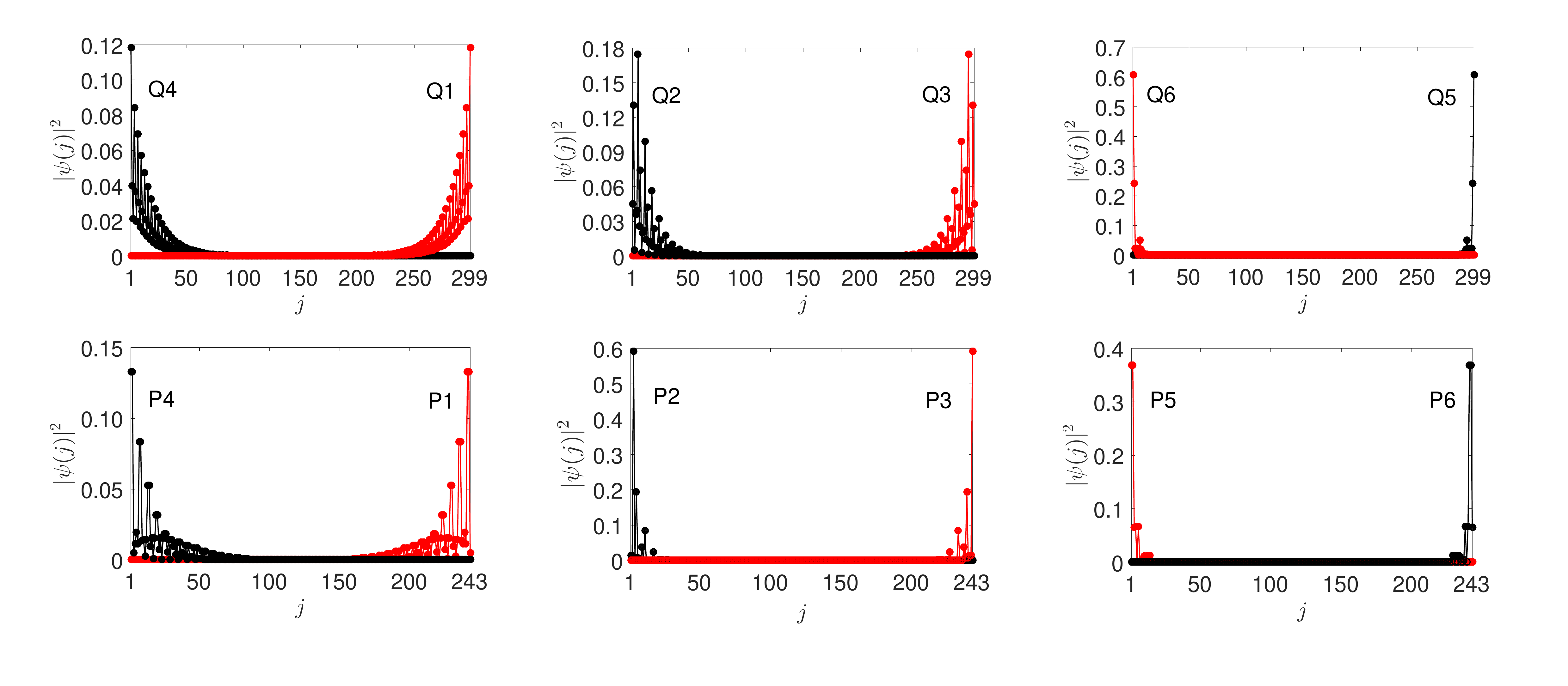}\\
  \caption{Spatial density distributions of the edge modes. Top panel: the zigzag-edge case. The result shows that, Q5 and Q6 are
  a pair of chiral edge modes at $E_{f}=1.80$ for 1/3 filling; Q1 and Q4, Q2 and Q3 are two pairs of chiral
  edge modes at $E_{f}=3.60$ for 2/3 filling. The label $j$ is the site index of the periodic repeating unit shown
  in Fig.~\ref{f5}(a), as an artificial chain with the order of B-R-G-$\cdots$-B-R.
  Bottom panel: the armchair-edge case. Intuitively, P5 and P6 are a pair of chiral edge modes at $E_{f}=1.535$ for the 1/3 filling;
  P1 and P4, P2 and P3 are two pairs of chiral edge modes at $E_{f}=3.578$ for the 2/3 filling. The label $j$ denotes
  the site index of the periodic repeating unit shown in Fig.~\ref{f5}(b), as an artificial chain with the
  order of R-B-G-R-G-B-$\cdots$-R-G-B-R-B-G. Particularly, the distributions of the edge modes with
  positive group velocity are shown in red, whereas black for negative one.}\label{f7}
 \end{figure*}

 As the principle of the bulk-edge correspondence tells \cite{principle_of_bulk_edge}, Chern number is closely related with the
 chiral edge modes. By choosing the same parameters used for edge-state spectra, we plot the spatial density distributions of
 the edge modes of Q1-Q6 in the zigzag-edge case and the edge modes of P1-P6 in the armchair one, in Fig.~\ref{f7}.
 We first analyze the zigzag-edge case. As the distributions show (up panels in Fig.~\ref{f7}), Q5 and Q6 are a pair of chiral
 edge modes at $E_{f}=1.80$ for 1/3 filling; Q1 and Q4, Q2 and Q3 are two pairs of chiral edge modes at $E_{f}=3.60$
 for 2/3 filling. In particular, the spatial density distributions of the edge modes with positive group velocity
 are shown in red, while black for those with the negative group velocity. We analyze the Chern number by means of the edge modes localized
 at the site of $j=N_{zigzag}$. From the above calculation, $C_{\frac{1}{3}}=1$, that is, the edge mode Q5 carries $C=1$. Naturally, we know that the edge
 modes Q3 and Q1 with the opposite group velocity both carry $C=-1$. Accordingly, we get $C_{1}=1$ and $C_{2}=-1+(-1)-1=-3$.

 For the armchair-edge case, there are also six edge modes marked as P1-P6, in which P5 and P6 are a pairs of chiral edge modes at
 $E_{f}=1.535$ (magenta dot-dashed line), P1, P4 and P2, P3 are another two pairs of chiral edge modes at $E_{f}=3.578$
 (green dashed line). We notice that the edge mode P6
 localized at the site of $j=N_{armchair}$ the armchair case has the same sign of the group velocity as the edge mode Q5 of
 the zigzag case. Hence, P6 carries $C=1$, which leads to $C_{1}=1$. Meanwhile, the edge modes P1 and P3 with positive group
 velocity both carry $C=-1$, thus $C_{2}=-1+(-1)-1=-3$. All these analyses are self-consistent with the phase diagrams in Fig.~\ref{f4}.

\section{\label{S4}Conclusion}

 In conclusion, we have studied the three-band dice model which is composed of three types of sublattices. Firstly, we investigated
 the dispersion relations of the energy bands, and found that the system has the metallic and bulk insulating phase.
 The metal-insulator phase diagrams were plotted in the $\Delta$-$\phi$ parameter space. Further, we evaluated the Chern numbers
 through the energy band theory in the bulk insulating phase. Interestingly, high Hall plateau was uncovered with $C_{\frac{1}{3}}=\pm 2~(C_{\frac{2}{3}}=\pm 2)$.
The quantizations of the Hall conductance were readily seen in the edge-state spectra. Finally, we verified the Chern
 numbers through the spatial density distributions of the edge modes according to the principle of the bulk-edge correspondence.

 Although the similar lattice structure has been studies in several electronic materials \cite{mater_1,mater_2,mater_3,mater_4,mater_5},
 due to the high tunability of the parameters in the cold-atom experiments, it is now possible for experimentalists to research
 topological phases of neutral atoms which never appear in these aforementioned researches. Thus, we hope that the system with high Hall plateau proposed here can be
 realized in the near-future cold-atom experiment.

\section{Acknowledgments}
The authors acknowledge support from the NSFC under
Grants No. 11835011 and No. 11774316.

\end{document}